\documentclass[12pt,preprint]{aastex}

\shorttitle{Multiperiodicity in $\delta\,$Ceti from MOST}
\shortauthors{Aerts et al.}

\begin{document}

\title{$\delta$~Ceti is not monoperiodic: seismic modeling of a $\beta$~Cephei
star from MOST\altaffilmark{1} spacebased photometry}

\author{C.\ Aerts\altaffilmark{2,3}, S.V.\ Marchenko\altaffilmark{4}, J.M.\
Matthews\altaffilmark{5}, R.\ Kuschnig\altaffilmark{5}, D.B.\
Guenther\altaffilmark{6}, A.F.J.\ Moffat\altaffilmark{7}, S.M.\
Rucinski\altaffilmark{8}, D.\ Sasselov\altaffilmark{9}, G.A.H.\
Walker\altaffilmark{10}, and W.W.\ Weiss\altaffilmark{11}}

\altaffiltext{1}{
Based on data from the MOST satellite, a Canadian Space Agency
mission, jointly operated by Dynacon Inc., the University of Toronto Institute
for Aerospace Studies and the University of British Columbia, with the
assistance of the University of Vienna.}

\altaffiltext{2}{Institute of Astronomy, University of Leuven, Celestijnenlaan
  200 B, B-3001 Leuven, Belgium, email: conny@ster.kuleuven.be}

\altaffiltext{3}{Department of Astrophysics, Radboud University
  Nijmegen, P.O. Box  9010, 6500 GL Nijmegen, the Netherlands}

\altaffiltext{4}{Department of Physics and Astronomy, Western Kentucky 
University
1906 College Heights Blvd 11077, Bowling Green, KY 42101-1077, USA}

\altaffiltext{5}{Department of Physics and Astronomy, University of British
Columbia, 6224 Agricultural Road, Vancouver, BC V6T 1Z1, Canada}

\altaffiltext{6}{ Department of Astronomy and Physics, St. Mary's University,
Halifax, NS B3H 3C3, Canada}

\altaffiltext{7}{D\'epartement de Physique, Universit\'e de Montr\'eal,
C.P. 6128, Succursale Centre-Ville, Montr\'eal, QC H3C 3J7, Canada}

\altaffiltext{8}{David Dunlap Observatory, University of Toronto, P.O. Box 360,
Richmond Hill, ON L4C 4Y6, Canada}

\altaffiltext{9}{Harvard-Smithsonian Center for Astrophysics, 60 Garden Street,
  Cambridge, MA 02138, USA}

\altaffiltext{10}{1234 Hewlett Place, Victoria, BC  V8S 4P7, Canada}

\altaffiltext{11}{Institut f\"ur Astronomie, Universit\"at Wien,
T\"urkenschanzstrasse 17, A-1180 Wien, Austria}

\begin{abstract}
The $\beta$~Cephei star $\delta$~Ceti was considered one of the few monoperiodic
variables in the class.  Despite (or perhaps because of) its apparently simple
oscillation spectrum, it has been challenging and controversial to identify this
star's pulsation mode and constrain its physical parameters seismically.
Broadband time-resolved photometry of $\delta$\,Ceti spanning 18.7 days with a
duty cycle of about 65\% obtained by the MOST (Microvariability \& Oscillations
of STars) satellite -- the first scientific observations ever obtained by MOST
-- reveals that the star is actually multiperiodic. Besides the well-known
dominant frequency of $f_1 = 6.205886$ d$^{-1}$, we have discovered in the MOST
data its first harmonic $2f_1$ and three other frequencies ($f_2 = 3.737$
d$^{-1}$, $f_3 = 3.673$ d$^{-1}$ and $f_4 = 0.318$ d$^{-1}$), all detected with
$S/N > 4$.  In retrospect, $f_2$ was also present in archival spectral line
profile data but at lower $S/N$. We present seismic models whose modes match
exactly the frequencies $f_1$ and $f_2$.  Only one model falls within the common
part of the error boxes of the star's observed surface gravity and effective
temperature from photometry and spectroscopy. In this model, $f_1$ is the radial
($\ell = 0$) first overtone and $f_2$ is the $g_2$ ($\ell = 2$, $m = 0$) mode.
This model has a mass of $10.2\pm 0.2 M_{\odot}$ and an age of 17.9$\pm$0.3
million years, making $\delta$~Ceti an evolved $\beta$~Cephei star.  If $f_2$
and $f_3$ are rotationally split components of the same $g_2$ mode, then the
star's equatorial rotation velocity is either 27.6 km~s$^{-1}$ or half this
value.  Given its $v~{\rm sin}~i$ of about 1 km~s$^{-1}$, this implies we are
seeing $\delta$~Ceti nearly pole-on.\end{abstract}

\keywords{stars: early-type; stars: individual (HD\,16582); stars: oscillations;
stars: $\beta\,$Cephei}

\section{Introduction}

Asteroseismic modeling of $\beta$~Cephei pulsators offers an important window on
the structure and evolution of massive evolved B stars which are precursors to
core-collapse supernovae. Significant progress has been made recently thanks to
groundbased single-site photometric campaigns lasting many years and multi-site
synoptic campaigns lasting several months.  The first method made possible the
identification of 6 pulsation modes in HD 129929 (B3V), leading to the first
observational proof of non-rigid rotation inside a star other than the Sun
(Aerts et al.\ 2003).  The second method resulted in the identification of about
20 frequencies in the star $\nu$ Eri (B2III, Handler et al.\ 2004, Aerts et al.\
2004, De Ridder et al.\ 2004) and a second strong case for differential internal
rotation (Pamyatnykh et al.\ 2004) in a massive B star.  These successes have
prompted similar studies of other $\beta$~Cephei stars; e.g., $\theta$ Oph
(B2IV, Handler et al.\ 2005; Briquet et al.\ 2005) and 12 Lac (B2III, Handler et
al.\ 2006). Photometry from a spacebased platform offers the advantages of both
methods, producing long time series of much higher duty cycles than are possible
from the ground.  This paper is an indication of what is possible when such data
become available for even a seemingly simple $\beta$~Cephei star which has been
investigated for decades with only limited progress.

The bright star $\delta\,$Ceti ($m_V=4.07$, B2IV) is one of the very few
$\beta\,$Cephei stars thought to be a monoperiodic pulsator within that class
(Stankov \& Handler 2005 and references therein). Its variability has been
investigated in a number of ground-based studies, most of which were based on
only a few nights of data. A clear overview of these studies up to 1987 is not
repeated here since it can be found already in Jerzykiewicz et al.\ (1988), who
investigated the star's behaviour from multicolor photometry taken during seven
consecutive nights in 1981 and one night in 1982, as well as from archival
data. These authors put forward one oscillation frequency equal to
6.20587545 d$^{-1}$ with an amplitude of $\approx\,12$\,mmag from data
assembled between 1963 and 1982.  From their new data, they found night-to-night
variations of the mean brightness and of the amplitude in the $uvby$
filters. They attributed the amplitude variability to either a secondary short
period with an amplitude below 1.6\,mmag or a slow drift in the data, or
both. They did not find any night-to-night phase variations.

Kubiak \& Seggewiss (1990) collected 2 nights of simultaneous spectroscopic and
photoelectric observations of the star and confirmed the phase lag between the
radial velocity and light curves of 0.23 found in earlier studies. 

Aerts et al.\ (1992) discovered large-amplitude line-profile variations in
$\delta\,$Ceti, from which they identified the single frequency as a radial mode
with a velocity amplitude of $7.4 \pm 0.1$ km~s$^{-1}$.  
This identification was confirmed from available multicolor photometry by
Cugier et al.\ (1994) and by Cugier \& Nowak (1997).

Finally, Daszy\'nska-Daszkiewicz et al.\ (2005) tried to put constraints on the
internal physics of the star and to identify the radial order of the mode from
combined multicolor photometry and radial-velocity data.  They found
significant differences between the data and their seismic models depending on
whether they adopted OPAL or OP opacities, and could not conclude definitely if
$\delta$~Ceti was pulsating in the radial fundamental or first overtone mode.
Neither model scenario could be made to agree with the observations.

New space-based observations of $\delta$~Ceti by the MOST satellite have now
provided key clues to understanding this star.  The first ground-based
observations of $\delta$~Ceti were obtained over a century ago, and over the
course of the last three to four decades, there has been a prolonged debate
about its amplitude modulation, possible multiperiodicity and period changes.
In that time, a grand total of about 30 cycles of its dominant pulsation mode
have been monitored. In less than three weeks, MOST was able to thoroughly
sample over 70 cycles.

\section{Observations and data reduction}

The MOST (Microvariability \& Oscillations of STars) satellite (Walker, Matthews
et al.\ 2003), housing a 15-cm Rumak-Maksutov optical telescope feeding a CCD
photometer, was launched in June 2003.  Its primary mission is ultraprecise
rapid photometry for asteroseismology.  MOST was designed to monitor stars at a
rate as high as 10 times per minute with a single-point precision of about 1-2
millimag.  For stars within the satellite's Continous Viewing Zone (CVZ) -- in a
declination range $-18^{\circ} \leq \delta \leq +36^{\circ}$ -- data can be
collected nearly without interruption for up to 8 weeks.  The combination of
single-point precision, and long time coverage with high duty cycle, leads to
sensitivity to rapid oscillations in Fourier space as low as about 1 $\mu$mag =
1 part per million (ppm).  An example of the photometric precision can be seen
in Matthews et al.\ (2004).

During the early stages of the mission when MOST was being checked out for
routine scientific operations, the MOST Team decided to use $\delta$~Ceti
(conveniently located in the CVZ) as its first test target, for what was
designated Commissioning Science.  At the time of the $\delta$~Ceti observations
-- the first scientific data collected by MOST -- the spacecraft pointing had
not yet been optimised, debugging of onboard software led to computer crashes
which introduced gaps into the time series, and the downlink to Earth did not
yet permit the nominal science data sampling rate. Hence, the photometric
precision and duty cycle were far from what MOST was capable of achieving once
commissioning was complete.  However, at the time, this light curve represented
the best combination of precision, duration and duty cycle ever obtained for any
astronomical object other than the Sun.

MOST monitored $\delta$~Ceti for 18.68 days during 08 - 27 October 2003, with
10-s exposures at a sampling interval of 120~s.  The data were obtained through
a broadband filter designed expressly for the MOST instrument, with a bandpass
of about $350 - 700$ nm and a throughput approximately $3\times$ that of a
Johnson V filter.  The data were reduced independently by two of the authors (SM
and RK), yielding similar results.  We present here the SM reduction, described
below.

The principal observing mode for MOST is known as Fabry Imaging, in which the
light from a bright star illuminates a Fabry microlens which projects an image
of the telescope pupil onto the CCD (see Walker, Matthews et al.\ 2003, and
their Fig.\ 10). The star is centred in a field stop 1 arcmin in diameter, and
the doughnut-shaped pupil image covers about 1300 pixels in a square CCD
subraster of 58 pixels on a side.  Seven adjacent Fabry microlenses sample the
surrounding sky background.  (One neighbouring lens is not used because its
reading is contaminated by light from the star as pixels are transferred under
the Target Star beam during readout.)

Data are returned to Earth in two formats, or Science Data Streams, known as
SDS1 and SDS2.  The former is processed on board so that only a small set of
integrated numbers is sent to Earth; the latter consists of resolved Fabry
images which can be fully reduced on the ground.  Because of the limitations of
the commissioning performance, the quality of the SDS1 data (numbering about
10,000) were severely compromised and are not used for this reduction.  The
Fabry images of the $\delta$~Ceti SDS2 data were binned $2 \times 2$ to form a
$29 \times 29$ pixel image.  For each image, all the available estimates of
biases were iteratively checked for significant outliers and, using appropriate
weights, combined to provide an average for a given image.

Most of the data reduction effort is made to correct for the changing
background, primarily due to orbital modulation of scattered Earthshine.
Detailed aspects of this stray light are described by Reegen et al.\
(2005). Readings from the 7 adjacent Fabry microlenses (after $3\sigma$-clipping
of outliers) were used to derive unweighted averages of the background. The
Primary Target Fabry field also contains about 350 pixels with only background
that were used to calculate a third background average.  This last estimate of
the background turned out to give the best removal of stray light artifacts, and
it alone was used for the current reduction.

Since $\delta$~Ceti was observed during satellite commissioning, and the
pointing accuracy was still far from optimal, the pointing errors (with
$\sigma_{X,Y} \sim 1$ pixel) were tracked for each exposure, based on the ACS
(Attitude Control System) telemetry.  The pointing errors (and other possible
inhomogeneities in the optics) were also estimated directly from the images
themselves by comparing the average fluxes in the four image quadrants. Large
deviations from quadrant to quadrant were flagged and could lead to rejection of
an image.

After bias and background subtraction, each Fabry target image was assigned a
quality rank. This was calculated by adding `penalties' and rejecting all images
with a penalty count exceeding 3. Any large deviation ($>3.5\sigma$ above
average) in the X or Y pointing of the satellite (updated once per second) added
1 to the penalty count. A similarly large deviation of the $(X^2+Y^2)$ pointing
vector, as well as an excessive number of $>3.5\sigma$ shifts from the average
position in a 10-sec exposure (more than 3 such deviations in a set of 10
successive readings) added 1.5 each to the penalty. Any large systematic shift
($>5$ pixels, either in X or Y) of the 10-sec average from the global average
position from the entire run resulted in immediate rejection of the image.
The rest of the quality ranking was based on dividing each Fabry image into four
quadrants and assesing the uniformity across these quadrants.  An iterative
calculation of the percentage of outlying pixels in each quadrant (usually due
to cosmic ray strikes or hot pixels in the CCD) sets the next component of the
rank.  If the number of rejected pixels exceeded 5\% or 10\%, the penalty was
increased by 1 or 2, respectively. An image was rejected if any single quadrant
contained more than 50\% of the total number of faulty pixels (even if that
number was below 10\%), or any combination of 2 quadrants was responsible for
75\% of the bad pixels.

The pre-selected pixels in each image which passed the quality control were
added to produce an instrumental flux value. The instrument telemetry also
includes the CCD focal plane temperature, which was not completely regulated
during the commissioning observations. Correlations between CCD temperature and
the image fluxes were corrected. The final stage of the reduction was high-pass
filtering and $3\sigma$-clipping of the data to remove any rapid changes in
stray light that survived the earlier quality tests.  For the period range of
relevance for $\delta$~Ceti, this is not expected to remove any intrinsic
stellar signal.  As a reality check, we always compared the corrected fluxes to
the original raw estimates.  Any large deviations ($\delta m \ge 0.005$ mag)
were re-examined to avoid the danger of spurious overcorrection.

The duty cycle of the original SDS1 + SDS2 photometry was about 65\%, with a
stretch of 11 days reaching about 95\%.  The exclusion of the SDS1 data, and the
data rejection above, reduces the total number of data points from 3267
to 2949. The
absolute duty cycle of this filtered light curve is only 22\%, but it should be
noted the reduction in duty cycle is primarily in each orbital cycle of 101.4
min, with no regular daily gaps as in groundbased photometry.  The overall
sampling of the $\delta$ Ceti light curve still approaches a duty cycle of 65\%.

\begin{figure}
\begin{center}
\rotatebox{270}{\resizebox{8cm}{!}{\includegraphics{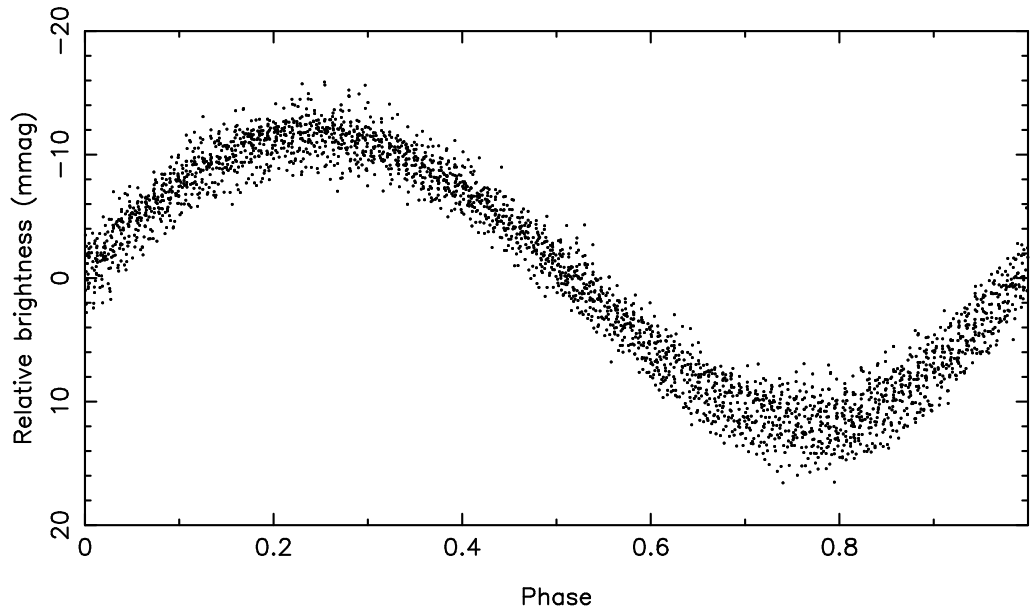}}}
\end{center}
\caption[]{Phase diagram of the MOST data of $\delta\,$Ceti folded
  at the known dominant pulsation period of 0.161138 days}
\label{fig0}
\end{figure}

The total standard deviation of the 2949 data point set is 8.47 mmag, which is
dominated by the intrinsic pulsational variability of the star.  The phase
diagram of the data, folded at the known dominant pulsation period, is presented
in Fig.\,\ref{fig0}. This plot, and the light curve segments shown in
Fig.\,\ref{fig5}, illustrate the thorough coverage and quality of the MOST
photometry of $\delta$~Ceti.

\section{Data analysis}

\subsection{Discovery of the multiperiodicity of $\delta$~Ceti}

\begin{figure}
\begin{center}
\rotatebox{270}{\resizebox{12cm}{!}{\includegraphics{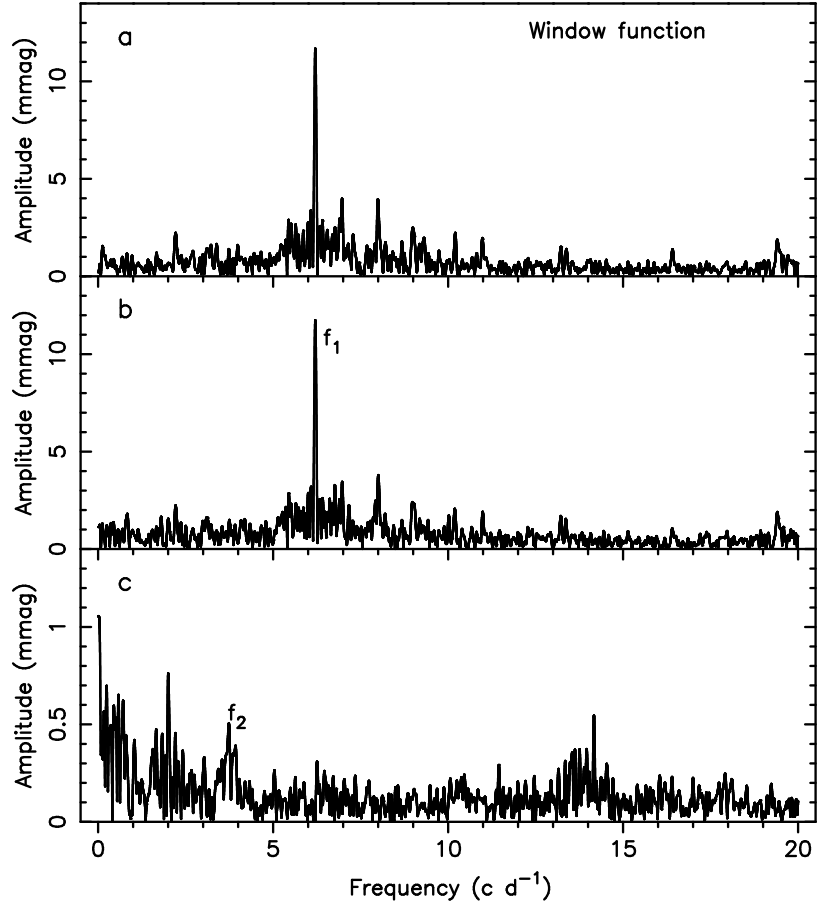}}}
\end{center}
\caption[]{Periodograms of the MOST photometry of $\delta\,$Ceti. Panel a:
  window function shifted and scaled to the main peak; panel b: periodogram
  of the data; panel c: periodogram after prewhitening with $f_1$ and
  $2f_1$.}
\label{fig1}
\end{figure}

We computed the Scargle Fourier periodogram (Scargle 1982) of the data
(Fig.\,\ref{fig1}b) with a sampling of $10^{-6}$ d$^{-1}$ and found the expected
dominant peak at $f_1 = 6.20589(8)$ d$^{-1}$ (= 71.827 $\mu$Hz); P = 0.16114 d =
3.867 hr.  The error estimate of 8$\times 10^{-5}$ d$^{-1}$ was computed as $
\sigma_f = \sqrt{6}\ \sigma_{\rm std} / \pi \sqrt{N}\ A\ \triangle T$
(Montgomery \& O'Donohue \& 1999), with $N$ the number of measurements, $A$ the
amplitude, $\triangle T$ the total time span and $\sigma_{\rm std}$ the standard
deviation of the final residuals.  This value of $f_1$ is identical to within
the errors, to the value reported by Jerzykiewicz et al.\ (1988). We then
compared the periodogram to the spectral window function (Fig.\ref{fig1}a),
which was computed from a sinusoid of the same frequency and amplitude as peak
$f_1$ sampled at the same times as the real time series.

A least-squares harmonic fit to the data, fixing the value of $f_1$ above, gives
an amplitude for that peak of 11.62(3) mmag, and reveals the presence of the
first harmonic $2f_1$ with an amplitude of 0.72(3) mmag and a signal-to-noise of
about 7.5$\sigma$.  While harmonics of principal oscillation frequencies have
been found in other $\beta$~Cephei stars (e.g., Heynderickx et al.\ 1994), one
was never before detected in $\delta$~Ceti.  This is not surprising given its
amplitude is only 6\% of that of $f_1$.

Subtracting $f_1$ and $2f_1$ from the data results in the periodogram shown in
Fig.\,\ref{fig1}c.  The overall variance of the time series has been reduced by
96.6\%, so the residuals have a standard deviation of only $\sigma = 1.56$ mmag.
Four peaks are immediately obvious in Fig.\,\ref{fig1}c: 0.02, 2.00, 3.73, and
14.20 d$^{-1}$.  The last corresponds to the known orbital frequency of the MOST
satellite, and is an observed artifact in later MOST data, due to the modulation
of scattered Earthshine in the MOST focal plane with the orbital period.  The
2.00 d$^{-1}$ frequency is also almost certainly an artifact.  It is due to a
modulation of the stray light from the Earth as MOST's Sun-synchronous dusk-dawn
orbit takes it over similar features of the Earth's albedo on a daily cycle.  In
this case, it may be due to the maxima in the Earth's albedo near both poles,
not long after the Autumnal Equinox.

The low-frequency peak near 0.02 d$^{-1}$ is consistent with a long-term
trend in the data, discussed below.  The remaining peak, labeled $f_2$
in Fig.\,\ref{fig1}c, cannot be ascribed to any known instrumental artifact and
is almost certainly intrinsic to $\delta$~Ceti.

\subsection{Long-term variability}

In its Fabry Imaging mode, MOST is a non-differential experiment, with no
comparison star observed in a comparable fashion in the field.  However,
experience with 2 years of MOST data has shown that the instrument is remarkably
stable.  Because of the previous observation by Jerzykiewicz et al.\ (1988) of a
monotonic amplitude change over 7 consecutive nights, and the presence of a
low-frequency peak in Fig.\,\ref{fig1}c, we searched for evidence of such
amplitude changes before proceeding with frequency analysis of the MOST
photometry.

\begin{figure}
\begin{center}
\rotatebox{270}{\resizebox{6cm}{!}{\includegraphics{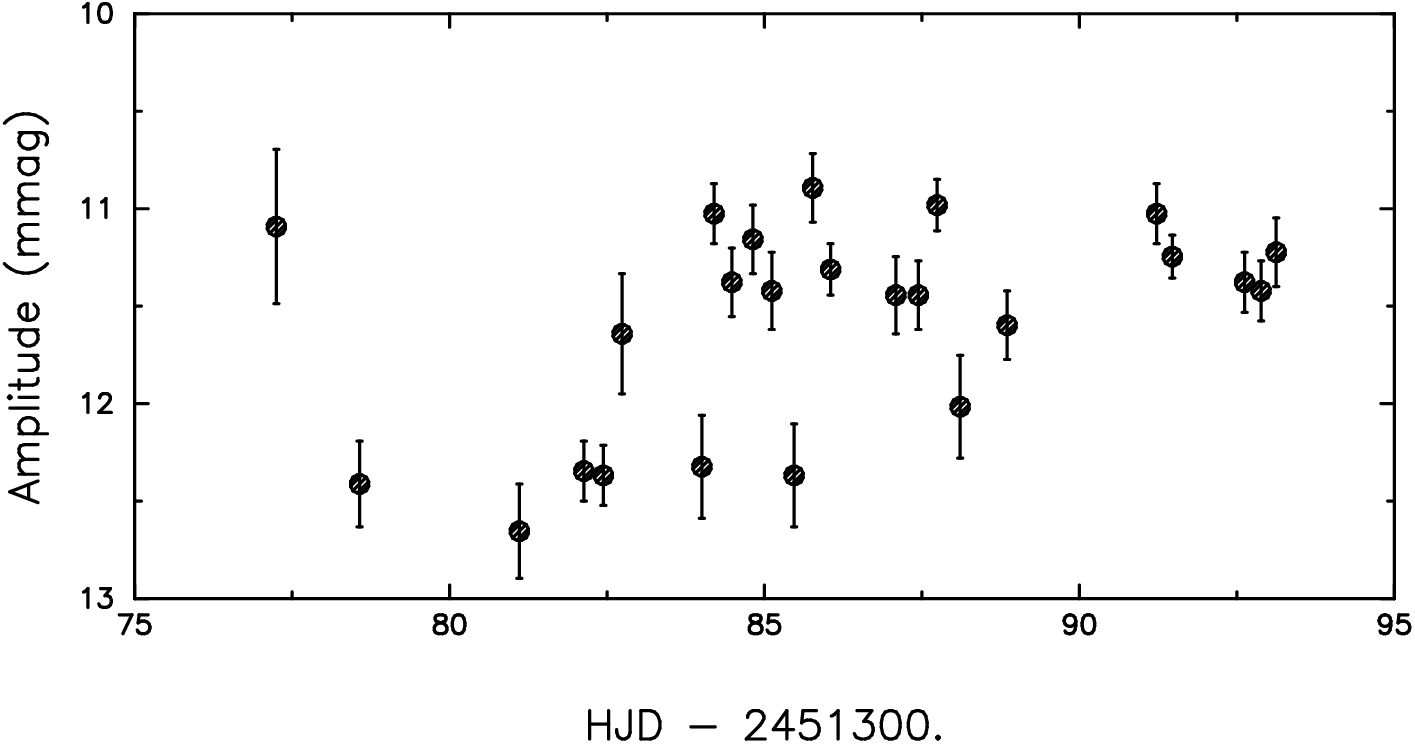}}}
\rotatebox{270}{\resizebox{6cm}{!}{\includegraphics{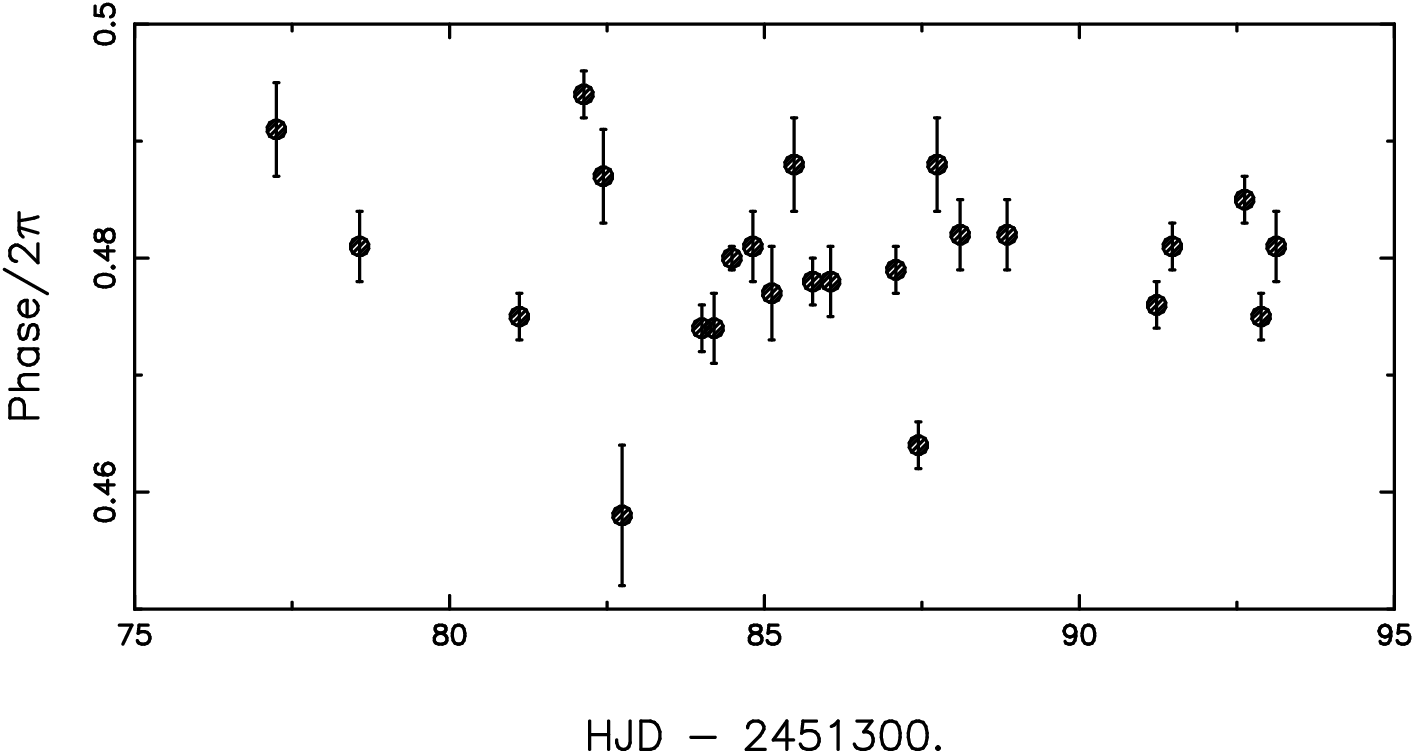}}}
\rotatebox{270}{\resizebox{6cm}{!}{\includegraphics{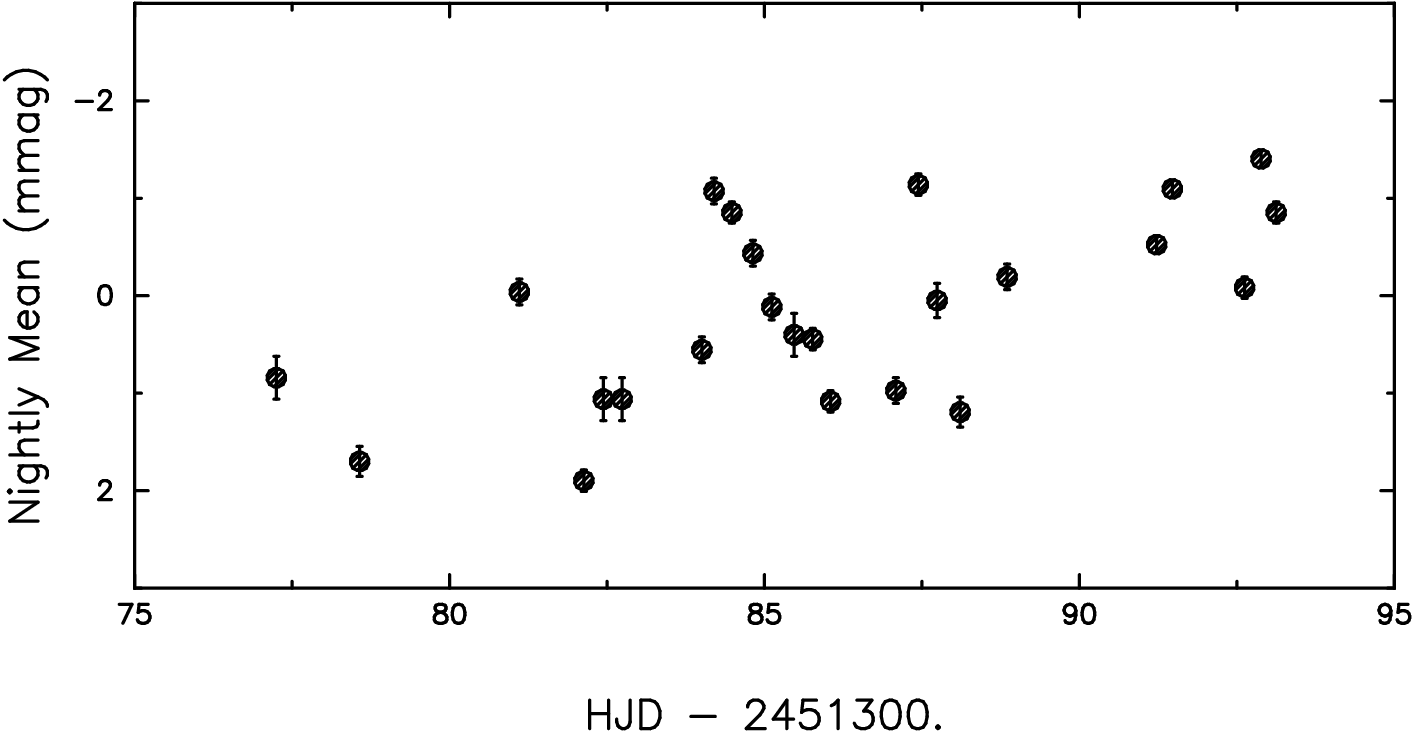}}}
\end{center}
\caption[]{Amplitude of $f_1$, phase of $f_1$ and mean with respect to
  the overall mean, for 25 datastrings observed by MOST.}
\label{fig2}
\end{figure}

We divided the time series into subsets.  Each subset was longer than the
dominant period of 0.17 d but shorter than 0.5 d so as to mimic the previous
ground-based data.  Each subset used had at least 50 data points and no large
gaps.  We produced 25 such subsets and fitted them with $f_1$ and $2f_1$,
allowing the amplitudes and phases to be free parameters.  The resulting
amplitudes, phases and ``nightly'' means of $f_1$ are plotted in
Fig.\,\ref{fig2}.

All three quantities vary beyond the error bars of the individual points, but in
a complicated fashion.  However, the total ranges of variability are small:
about 15\% in amplitude, 4\% in phase, and about 0.4\% in mean brightness.  The
mean brightness does show the clearest evidence for a trend, which turns out to
be responsible for the peak near 0.02 d$^{-1}$ (P $\sim$ 50 d).

\begin{figure}
\begin{center}
\rotatebox{270}{\resizebox{10cm}{!}{\includegraphics{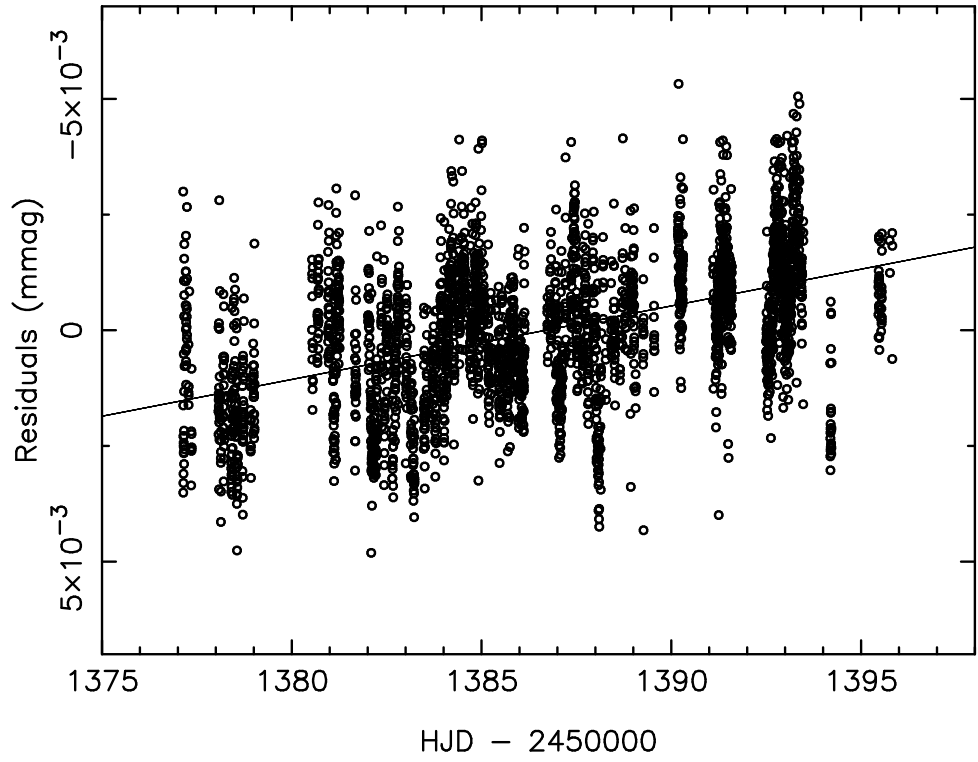}}}
\end{center}
\caption[]{Brightness increase of $\delta\,$Ceti in the residuals after
  prewhitening the dominant frequency $f_1$ and its harmonic from the MOST
  lightcurve.}
\label{fig3}
\end{figure}

We then fitted the data residuals from Sect.\,3.1 ($f_1$ and $2f_1$ prewhitened)
with a linear trend as shown in Fig.\,\ref{fig3}.  The linear fit corresponds to
a brightness increase of 0.154 mmag/day = 0.0064 mmag/hour.  This is more than 2
orders of magnitude smaller than the trend of about 1 mmag/hour reported by
Jerzykiewicz et al.\ (1988) in their Str\"omgren photometry of $\delta$~Ceti.

We cannot confirm whether the trend seen by MOST is stellar or instrumental (see
discussion in Sect.\,5) but we nonetheless remove it from the data for
subsequent analysis.  The detrended residuals have a standard deviation of 1.37
mmag, 0.19 mmag smaller than before.

\subsection{Additional oscillation frequencies in $\delta$~Ceti}

\begin{figure}
\begin{center}
\rotatebox{270}{\resizebox{12cm}{!}{\includegraphics{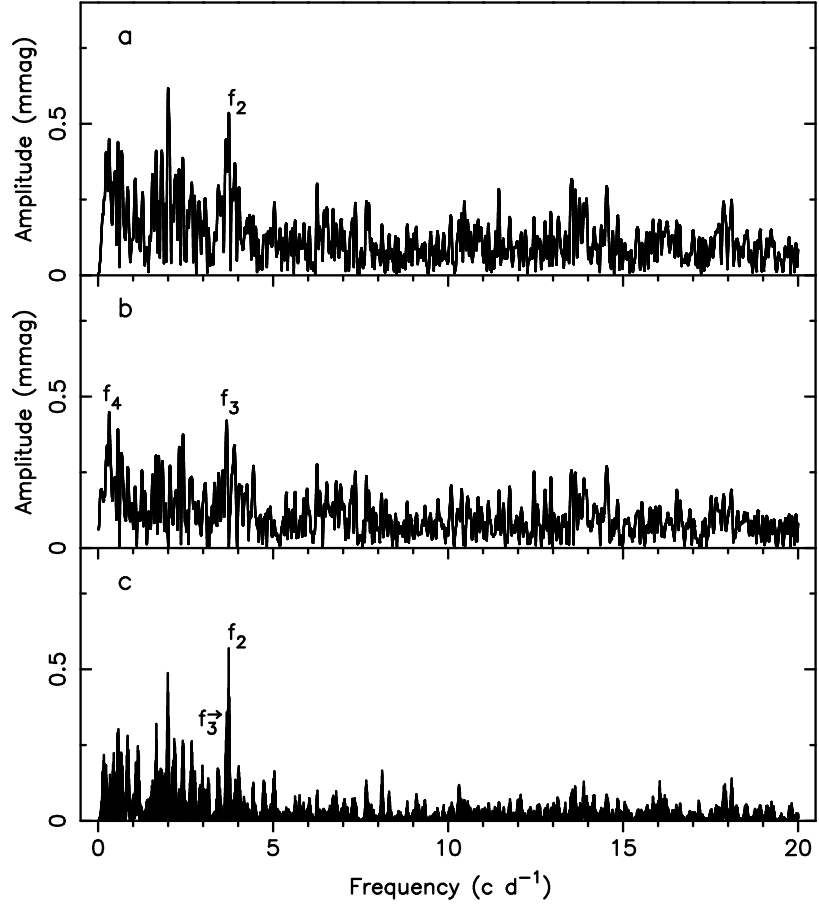}}}
\end{center}
\caption[]{Periodograms of $\delta\,$Ceti. Panel a: after prewhitening with
  $f_1$ and $2f_1$ and detrending according to Fig.\,\protect\ref{fig3}. Panel
  b: after subsequent prewhitening with $f_2$ and 2.003 d$^{-1}$. Panel c:
  product of normalised amplitude spectra of the MOST and HIPPARCOS photometry
  and the first velocity moment $<v>$ taken from Aerts et al.\ (1992).}
\label{fig4}
\end{figure}

\begin{figure}
\begin{center}
\rotatebox{270}{\resizebox{8cm}{!}{\includegraphics{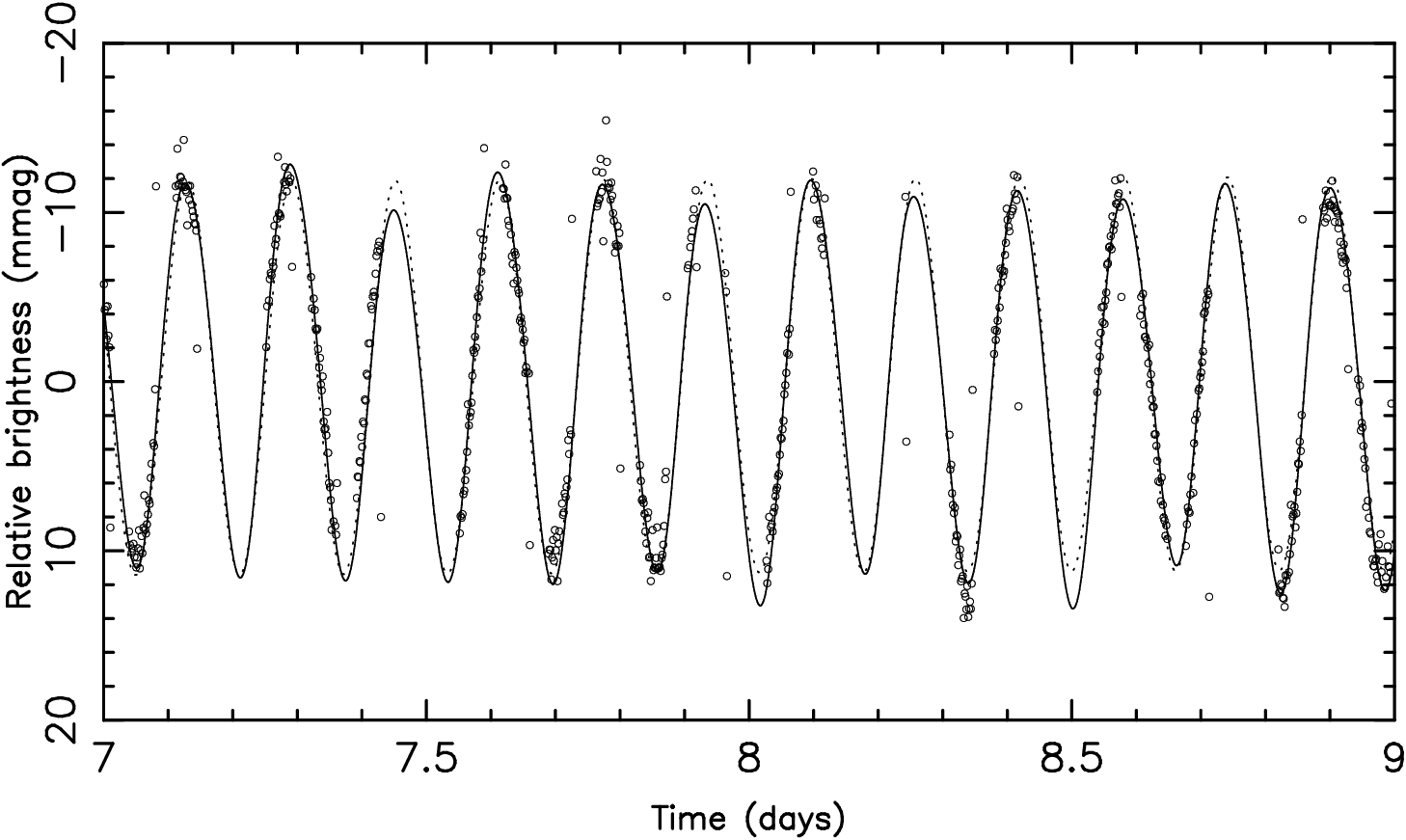}}}
\rotatebox{270}{\resizebox{8cm}{!}{\includegraphics{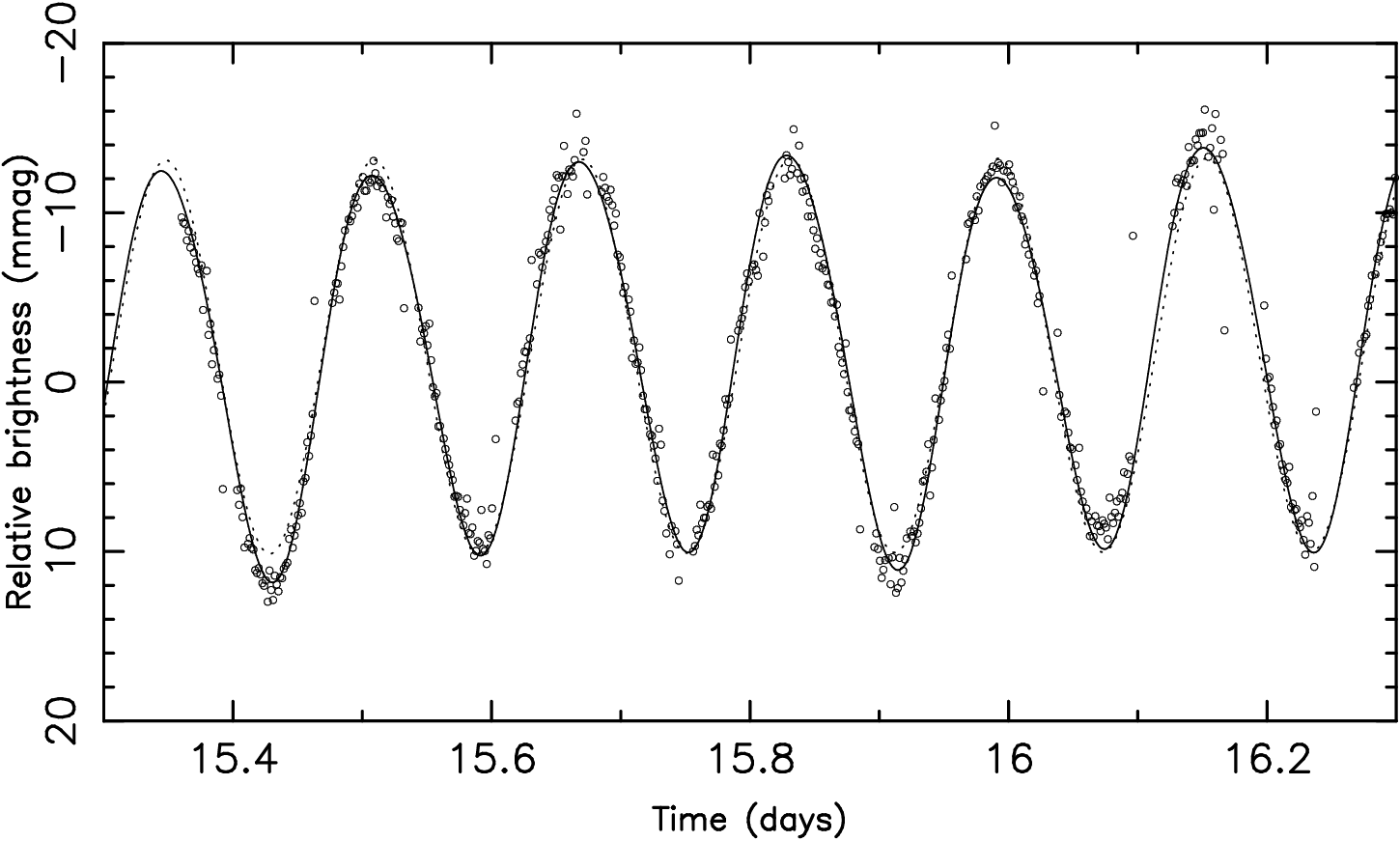}}}
\end{center}
\caption[]{Comparison between the MOST data and (1) the final fit given in
Eq.\,(1) (full line), (2) a fit including only the dominant frequency and the
trend (dotted line) for a few selected segments.}
\label{fig5}
\end{figure}

Removing the trend from the data presented in Fig.\,\ref{fig3} produced the
periodogram shown in Fig.\,\ref{fig4}a.  Not surprisingly, the peak at 0.02
d$^{-1}$ vanished.  Perhaps surprisingly, the peak at 14.2 d$^{-1}$ also
disappeared.  It turned out not to be due primarily to modulated stray
Earthshine but to the long-term trend sampled with gaps in the data at the
orbital frequency (due to outages during the spacecraft commissioning phase).

The peak at 2.003 d$^{-1}$ does remain, consistent with a genuine stray light
artifact, although its amplitude has been reduced by more than half, so the
trend contributed to it as well.

The frequency $f_2 = 3.737(2)$ d$^{-1}$ persists as well, and its amplitude of
0.53(3) mmag -- significant at the $5.6\sigma$ level -- is almost unchanged from
the original data before the trend was removed.  We note that this frequency and
the dominant frequency $f_1$ have a beat period $1/(f_1 - f_2) = 0.405$ d $\sim
9.7$ hr, which is well sampled by several continuous stretches in the MOST time
series but not by the nightly stretches in the groundbased data of Jerzykiewicz
et al.\ (1988).

Prewhitening the MOST residuals by frequency $f_2$ and the artifact at 2.003
d$^{-1}$ reduces the standard deviation by 0.12 mmag to 1.25 mmag.  From these
residuals, we obtain the periodogram plotted in Fig.\,\ref{fig4}b.  This
contains two peaks with significance greater than $4\sigma$, namely $f_3 =
3.673(2)$ d$^{-1}$ and $f_4 = 0.318(2)$ d$^{-1}$, with amplitudes of 0.39(4)
($4.0\sigma$) and 0.43(4) mmag ($4.5\sigma$), respectively.

\begin{figure}
\begin{center}
\rotatebox{270}{\resizebox{8cm}{!}{\includegraphics{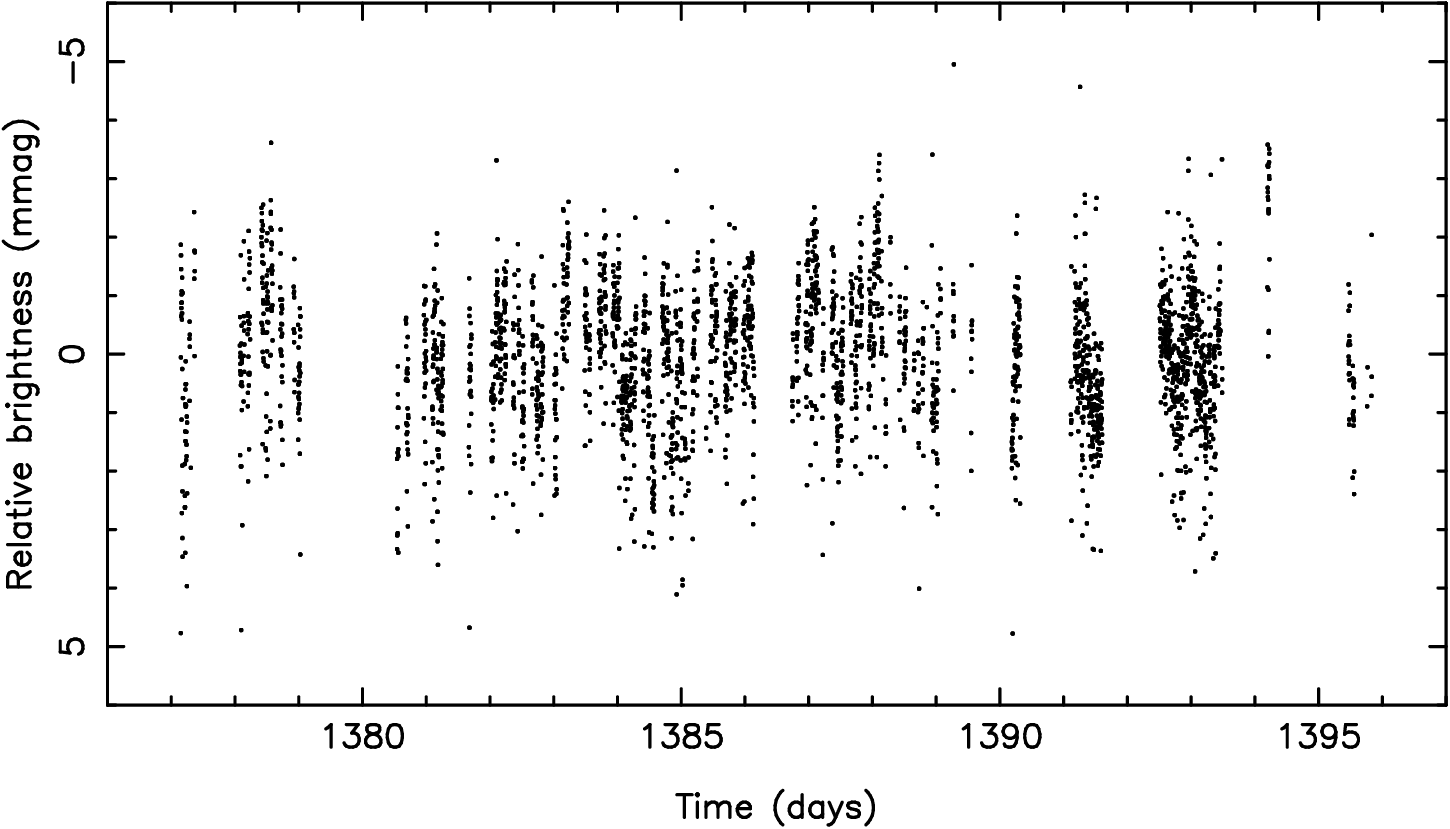}}}
\rotatebox{270}{\resizebox{8cm}{!}{\includegraphics{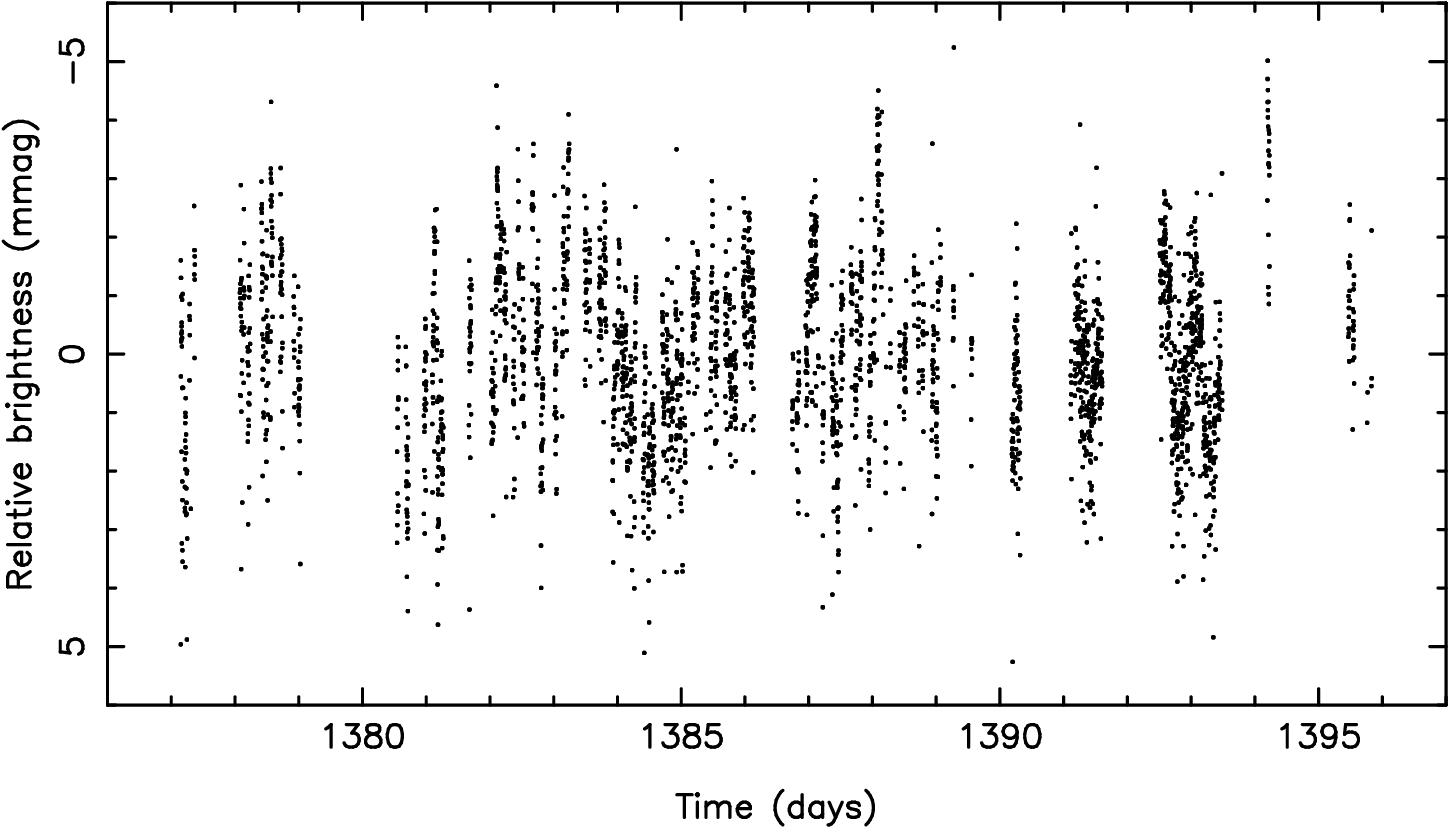}}}
\end{center}
\caption[]{Residuals of the entire MOST data set after
prewhitening by the final fit given in Eq.\,(1) (upper panel) and
by only the dominant frequency $f_1$ and the linear trend in the
data (lower panel).  The standard deviations of these residuals are
1.18 and 1.47 mmag, respectively.}
\label{fig8}
\end{figure}

Prewhitening the data by these two additional frequencies reduces the
overall $\sigma$ by only 0.07 mmag. A plot of the final residuals is
given in Fig.\,\ref{fig8} and their periodogram is shown in Fig.\,\ref{fig4}c.

There are peaks with amplitudes below 0.4 mmag which may help account for the
complex amplitude and phase behaviour seen in Fig.\,\ref{fig2}.  These include a
peak at 3.909(5) d$^{-1}$ at the $3.5\sigma$ level, and one at 3.805(6) d$^{-1}$
at the $2.8\sigma$ level.

\subsection{Reexamination of archival data}

In light of this frequency analysis of the MOST data, we reanalysed
two high-precision archival data sets of $\delta$~Ceti:
\begin{enumerate}
\item
the HIPPARCOS light curve (Perryman et al.\ 1997), consisting of 72 data points
   covering 1096 days with a quasi-equidistant spacing of about 15 days; and
\item
the 60 moment variations derived from single-site high-resolution
   line-profile observations by Aerts et al.\ (1992), spanning 7 days.
\end{enumerate}
Neither of these data sets showed any significant long-term trends.  After
prewhitening both independent data sets with frequency $f_1$, we searched for
evidence of $f_2$ in the residuals.  There was no sign of signal at $f_2$ in the
HIPPARCOS photometry, which is not surprising given its low amplitude in the
MOST data and the poor sampling of the HIPPARCOS data for this relatively short
period.  However, in a periodogram of the first velocity moment $<v>$ computed
by Aerts et al.\ (1992) from their line-profile data, a peak shows up near
$f_2$, with an amplitude of $0.27 \pm 0.09$ km~s$^{-1}$, corresponding to a
significance of only $3\sigma$.  Despite the low significance, and because of
its presence in the MOST data, we conducted a second investigation of $f_2$ in
the archival data.

The dominant frequency $f_1$ is present in the HIPPARCOS data with a comparable
amplitude to that seen in the MOST data, despite the use of different custom
filters for these observations.  So we prewhitened the HIPPARCOS and MOST
(detrended) lightcurves and the $<v>$ moment data by $f_1$ and $2f_1$, and then
normalised the resulting Scargle periodograms to the highest peak in each. We
then multiplied these normalised periodograms together, with the logic that if
$f_2$ (or any other common frequency) was present in more than one of the data
sets, it would show up with improved $S/N$ over the MOST data alone.  If $f_2$
was absent in the other sets, it would reduce the $S/N$ over MOST alone.  In
this exercise, all three data sets were given equal weight.

The outcome is shown in Fig.\,\ref{fig4}c.  Two peaks are immediately obvious to
the eye; at $f_2$ and 2.00 d$^{-1}$.  The latter has been reduced in
signficance, consistent with it being an artifact in the MOST data alone, while
$f_2$ now has a significantly higher $S/N$ of 8.6.  Closer examination (and
prewhitening) reveals the presence of a third peak at $f_3$ at the $4.6\sigma$
level, higher than in the MOST data alone.  The frequency $f_4$ is not evident
in the combined normalised periodogram.

These results lend circumstantial support to the presence of the frequencies
$f_2$ and $f_3$ in one or both of the HIPPARCOS and line-profile data sets, with
the {\em a posteriori\/} knowledge from the MOST photometry.

\subsection{Solution to the MOST light curve}

The final fit which we applied to the MOST light curve is as follows:
\begin{equation}
y_i = a + b t_i + \sum_{j=1}^6 c_j \sin [2\pi (f_j t_i + \phi_j)],
\label{vgl}
\end{equation}
with the parameters provided in Table\,\ref{table1}.  We have retained only
frequencies with amplitudes of significance $\geq 4.0\sigma$, following the
acceptance criterion proposed by Breger et al.\ (1993) and Kuschnig et al.\
(1997).

Two segments of the light curve comparing this solution to the data are shown in
Fig.\,\ref{fig5}.  The full line corresponds to Equation\,(\ref{vgl}), while the
dotted line is a solution including only the linear trend and the dominant
frequency $f_1$.  The residuals for the two solutions are shown in
Fig.\,\ref{fig8}, at a magnified vertical scale.  In particular, the full
solution does a much better job near the well-populated minima and maxima of the
light curve and leads to smaller residuals.

\section{Seismic interpretation}

\subsection{Fitting $f_1$ and $f_2$}

To compare the oscillation frequencies we have detected in $\delta$~Ceti with
those predicted by pulsational models of B-type stars, we explored the database
described by Ausseloos et al.\ (2004). This database contains evolutionary
models from the ZAMS to the turnoff with masses ranging from 7 to 13\,$M_\odot$
(in steps of $0.1 M_{\odot}$), $X$ fixed at 0.70, $Z$ ranging from 0.012 to
0.030 (in steps of 0.002), and three choices for the core overshoot parameter
($\alpha_{\rm ov} = 0.0$, 0.1 or 0.2).  For a description of the input physics
of these models, and the computation of their oscillation frequencies, we refer
to Ausseloos et al.\ (2004).

Since analyses of the line-profile variations and earlier multicolor photometry
indicate that the dominant pulsation frequency $f_1$ corresponds to a radial
mode (Aerts et al.\ 1992; Cugier et al.\ 1994), we have restricted our search of
the model grid to those models with radial modes which coincide with $f_1$.  We
have further constrained the search by forcing $f_2$ to correspond to zonal
modes of ${\ell}_2 = 0$, 1 or 2, using the fitting algorithms of Ausseloos
(2005).  Modes with $\ell \geq 3$ are unlikely to be observed in integrated
photometry due to cancellation effects across the stellar disk. The assumption
of non-zonal modes ($m \neq 0$) is considered safe since $\delta$~Ceti has a low
projected rotational velocity ($v$~sin~$i$ $\sim$ 1 km~s$^{-1}$; Aerts et al.\
1992) so even a non-zero $m$ component of mode $f_2$ should lie close to the
central component of the multiplet.  (See our discussion in Sect. 4.2 about the
inferred rotational velocity of $\delta$~Ceti and the validity of this
assumption.)

\begin{figure}
\begin{center}
\rotatebox{270}{\resizebox{12cm}{!}{\includegraphics{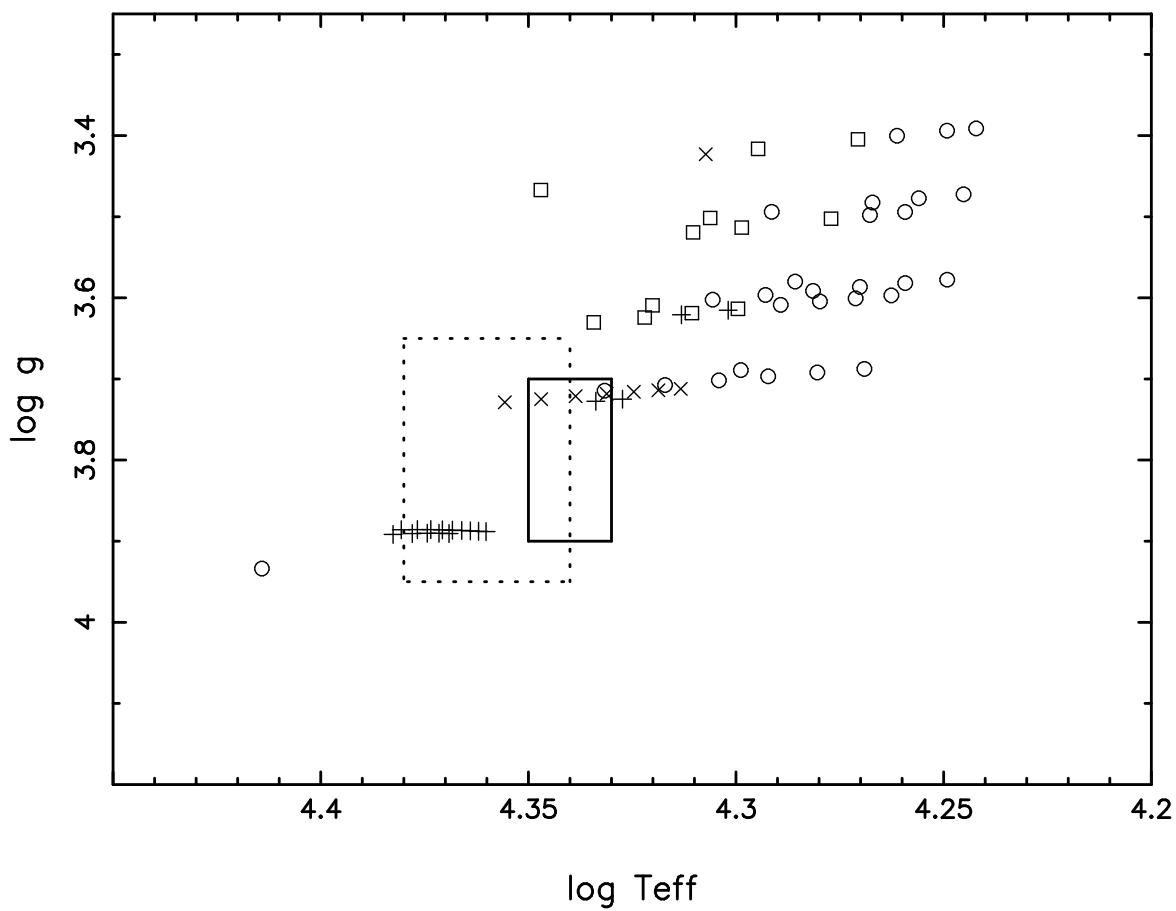}}}
\end{center}
\caption[]{Position of stellar models in the $(\log T_{\rm eff}, \log g)$
diagram whose zonal modes fit $f_1$ and $f_2$ simultaneously. The symbol
convention is as follows: $f_1$ radial first overtone and $f_2$ an $\ell=1,m=0$
$g$ mode ($\circ$), $f_1$ radial second overtone and $f_2$ an $\ell=1,m=0$ $g$
mode ($\Box$), $f_1$ radial fundamental and $f_2$ an $\ell=2,m=0$ $g$ mode
($+$), $f_1$ radial first overtone and $f_2$ an $\ell=2,m=0$ $g$ mode
($\times$).  The photometric (full lines) and spectroscopic (dotted lines)
observational error boxes of $\delta\,$Ceti are also indicated.}
\label{fig6}
\end{figure}

There are no models in the database which are consistent with both $f_1$ and
$f_2$ being radial ($\ell = 0$) modes.  We found 66 models which satisfy the
criteria that $f_1$ is a radial mode and $f_2$ is an $\ell = 1$ or $2$ mode; 40
models for the former, and 26 for the latter. All the models fitting $f_1$ and
$f_2$ simultaneously are indicated in the theoretical (log~$T_{\rm
eff}$,log~$g$) diagram shown in Fig.\,\ref{fig6}.

We also have independent empirical constraints on the position of $\delta$~Ceti
in Fig.\,\ref{fig6}.  We plot the star's empirical error box derived from
photometric colors averaged over the dominant pulsation cycle in three
different systems (Walraven, Geneva and Str\"omgren) as determined by
Heynderickx et al.\ (1994).  We also show the spectroscopic error box recently
derived by Morel et al.\ (private communication) based on high-resolution
\'echelle spectra covering the whole pulsation cycle.  We prefer this over
previous estimates relying on a single spectrum during the cycle (e.g., Gies \&
Lambert 1994).  We put more weight on the photometric error box due to its
smaller extent, and the fact that it is based on three different photometric
systems.

It can be seen in Fig.\,\ref{fig6} that the majority of the models matching both
$f_1$ and $f_2$ are too evolved (i.e., have too low gravities) for the
photometric and spectroscopic error boxes.  In fact, only five models fall
within the photometric box; their physical characteristics are listed in
Table\,\ref{table2}.  The non-radial mode corresponding to $f_2$ is the $g_2$
mode for all of these models.  All these models require some amount of core
overshooting.

Which of these models is most likely to be excited?  We checked the excitation
rates with the non-adiabatic code MAD (Dupret 2001) and found that the radial
first overtone of Model 1 in Table\,\ref{table2} (shown as ``o'' in
Fig.\,\ref{fig6}) is expected to be stable due to its low metallicity ($Z =
0.012$).  All eight modes of the other four models were found to be excited.

The high metallicity ($Z = 0.028$) of Model 2 (where $f_1$ is the radial
fundamental mode; ``+'' in Fig.\,\ref{fig6}) may also rule it out, since
evidence points to $\delta$~Ceti having a lower value of $Z$ than this.
Niemczura \& Daszy\'nska-Daszkiewicz (2005) recently derived $[m/H] = -0.24 \pm
0.09$ from UV iron transitions measured by IUE.  Morel et al.\ (private
communication) have derived the abundances of several important $Z$-determining
elements from optical \'echelle spectra and found them to be only slightly less
on average than solar values from Grevesse \& Sauval (1998); $Z = 0.02$.

We conclude therefore that the dominant mode of $\delta$~Ceti is the radial
first overtone.  Cugier et al.\ (1994) and Cugier \& Nowak (1997) reached the
same conclusion previously, but Daszy\'nska-Daszkiewicz et al.\ (2005) recently
cast doubt on this indentification, citing an excitation problem.  Our
non-adiabatic analysis shows that this mode is unstable for our Models 3, 4 and
5.  Of these options, we prefer Model 3 because: (1) it is situated in the
overlapping region of the photometric and spectroscopic error boxes, and (2) it
has a metallicity ($Z = 0.020$) consistent with values derived from high-quality
spectra (whereas the other two models appear to be too metal-rich).  The
frequency $f_2$ in this model is the $g_2$ $\ell = 2$ mode.

\subsection{Multiplet structure in the $\delta$~Ceti eigenspectrum?}

What of frequency $f_3 = 3.673$ d$^{-1}$?  We note that it is separated from
$f_2$ by $\Delta f = 0.064$ d$^{-1}$.  If $f_3$ is part of a rotationally-split
multiplet, then there should be another sidelobe frequency at $f_2 + \Delta f =
3.801$ d$^{-1}$.  Note that we did find a peak in the MOST data at a frequency
of 3.805 d$^{-1}$ but with a significance of only $2.8\sigma$.

If $f_2$ and $f_3$ are consecutive $m$-values ($\Delta m = 1$) of the same $\ell
= 2$ quintuplet, and we adopt the Ledoux coefficient $\beta_{-2,2}=0.85$ and
radius of Model 3, we obtain a rotational frequency of 0.075 d$^{-1}$ and an
equatorial rotation velocity of 27.6 km~s$^{-1}$.  If they are separated by
$\Delta m = 2$, then we derive half these values.

If we combine these values with the well measured value of $v$~sin~$i$ = 1$\pm$
1 km~s$^{-1}$ (Aerts et al.\ 1992), the inclination angle of $\delta$~Ceti may
be as small as $i = 2^{\circ}$ (and cannot be larger than $i = 8^{\circ}$).
Hence we must be observing the star nearly pole-on.  This conclusion is
independent of the choice of models in Table\,\ref{table2}.

The pole of a star corresponds to an angle of complete cancellation for an
$\ell=2$ sectoral and tesseral mode (Chadid et al.\ 1999, their Table\,A.1). It
is therefore most likely that $f_2$ corresponds to the central peak of the
quintuplet, as assumed in the modeling. Even if it were the outermost component
of the quintuplet, the central peak would differ only by $\simeq 0.13$ d$^{-1}$
from $f_2$.  Such a frequency shift is not large enough to affect our model
identifications which assumed a low rotation rate and hence, closely spaced
multiplet structure.

We also called attention in Sect.\,3.3 to another frequency, at 3.909 d$^{-1}$,
with a significance of $3.5\sigma$.  This frequency would not fit into an
equidistant quintuplet structure which includes $f_2$ and $f_3$.  We do note
that Model 3 has a $g_3$ $\ell = 3$ mode predicted to be excited whose
frequency is close to 3.909 d$^{-1}$.

Finally, we have no obvious explanation for frequency $f_4 = 0.318$ d$^{-1}$,
but we do point out that $f_4 \simeq 5(f_2 - f_3) \simeq 3(3.909 - 3.805)$
d$^{-1}$, within the errors.  Perhaps it is related to beating between modes.

\section{Summary}

MOST photometry of the $\beta$~Cephei star $\delta$~Ceti -- until now regarded
as a prototypical example of a monoperiodic radial oscillator within the class
-- reveals the presence of at least two additional oscillations consistent with
non-radial modes, as well as the first harmonic of the dominant radial mode.  We
have compared the new frequency spectrum with pulsation models constrained by
stellar parameters based on photometric colors and spectroscopic analysis.  We
conclude that the dominant mode in $\delta$~Ceti is due to the radial first
overtone, and that the next strongest mode is the $g_2$ $\ell = 2$ mode.  We
investigated multiplet structure associated with the latter mode to constrain
the rotational velocity of the star, and showed that $\delta$~Ceti is seen
nearly pole-on.

We also find a shallow linear brightness increase in the star, at a rate of
about 0.0064 mmag/hour.  Jerzykiewicz et al.\ (1988) also found a drift in their
$ubvy$ photometry of $\delta$~Ceti, but at a rate of about 1 mmag/hour.  If both
trends were truly intrinsic to the star, then the rate of brightness change
would be strongly variable as a function of epoch. Such variations would not be
surprising for an evolved star like $\delta$~Ceti undergoing small instabilities
on its way to the end of the central hydrogen burning phase.  However, it is
still possible that the trend seen by Jerzykiewicz et al.\ (1988) was an
uncorrected effect of extinction in the Earth's atmosphere. MOST does not suffer
such extinction effects in orbit, but as a non-differential experiment, it is
impossible to exclude an instrumental origin for the gradual trend it measured.

We note that Jerzykiewicz et al.\ (1988) also observed amplitude modulation in
their night-to-night photometry which they suggested might be due to a second
short-period variation with an amplitude below 1.6 mmag.  The detection by MOST
of frequency $f_2$, with an amplitude of about 0.5 mmag is consistent with that
explanation.

The best model fit indicates that $\delta$~Ceti has a mass of $M=10.2\pm 0.2
M_{\odot}$, an age of 17.9$\pm$0.3 million years and core overshooting
($\alpha_{\rm ov} = 0.20\pm 0.05$).  It is only the third $\beta$~Cephei star
for which the core overshooting parameter could be determined, along with
HD~129929 ($\alpha_{\rm ov} = 0.10\pm 0.05$, Aerts et al.\ 2003) and
$\nu$~Eridani ($\alpha_{\rm ov} < 0.13$, Pamyatnykh et al.\ 2004). The
determination of the core overshooting parameter results in an accurate seismic
mass estimate, which is for all three stars an improvement by an order of
magnitude over previous photometric or spectroscopic mass estimates.

These results illustrate the power of nearly continuous ultra-precise photometry
in understanding the structure and evolution of $\beta$~Cephei stars. We can
anticipate additional results from the space-based observatories WIRE (Br\"untt
et al., in preparation), MOST, COROT and Kepler in the coming years.

\acknowledgments

CA is indebted to Peter De Cat, Richard Scuflaire, Marc-Antoine Dupret and Mario
Ausseloos for the use of their software. CA is supported by the Research Council
of the K.U.Leuven under grant GOA/2003/04. SM acknowledges financial support
from the Kentucky Space Grant Consortium. JMM, DBG, AFJM, SR and GAHW
acknowledge funding from the Natural Sciences \& Engineering Research Council
(NSERC) Canada.  RK's work is supported by the Canadian Space Agency.

\clearpage

\clearpage

\clearpage

\clearpage

\clearpage

\clearpage

\clearpage

\clearpage

\clearpage

\begin{deluxetable}{ccccc}
\tablecaption{Final lightcurve solution of $\delta\,$Ceti, according to the
eight terms given in Eq.\,(\protect\ref{vgl}).  The reference epoch for the
phases $\phi_j$ is the time of the first measurement (HJD\,2451377.140428).}
\tablehead{\colhead{$f_j$ (c\,d$^{-1}$)} & \colhead{$f_j$ ($\mu$Hz)} &
\colhead{$c_j$ (mmag)} & \colhead{$\phi_j$} & \colhead{S/N}} \startdata
$f_1=6.20589(8)$ & 71.8274(9) & 11.62(3) & 0.4958(4) & 121.4\\ $2f_1$ & $2f_1$
& 0.72(3) & 0.619(7) & 7.5\\ 2.003(1) & 23.18(1) & 0.65(3) & 0.292(7) & 6.8\\
$f_2=3.737(2)$ & 43.25(2) & 0.53(3) & 0.454(9) & 5.5\\ $f_3=3.673(2)$ & 42.51(2)
& 0.39(4) & 0.82(1) & 4.0\\ $f_4=0.318(2)$ & 3.68(2) & 0.43(4) & 0.55(2) & 4.5\\
\tableline && $a=922.89(5)$ mmag & \\ && $b=-0.154(5)$ mmag/day & \\
\enddata\label{table1}
\end{deluxetable}

\clearpage

\begin{deluxetable}{ccccccccccc}
\tabcolsep=4.5pt
\tablecaption{Physical parameters of the stellar models within the photometric
  error box of $\delta\,$Ceti shown in Fig.\,\protect\ref{fig6}. The core
  overshoot parameter is expressed in units of the local pressure scale
  height.}
\tablehead{\colhead{} & 
\colhead{$\log\,L/L_{\odot}$} &
\colhead{$\log\,T_{\mbox{eff}}$} & 
\colhead{$\log g$} &
\colhead{$M(M_{\odot}$)} & 
\colhead{$R(R_\odot)$} & 
\colhead{$X_c$} &
\colhead{$X$} & 
\colhead{$Z$} & 
\colhead{$\alpha_{\rm ov}$} &
\colhead{age(yr)}}
\startdata
1:$\circ$ & 3.977 & 4.332 & 3.714 & 9.43 & 7.06 & 0.088 & 0.70 & 0.012 & 0.0 & 
19.7$\times 10^6$\\
2:$+$ & 4.014 & 4.334 & 3.727 & 10.36 & 7.29 & 0.252 & 0.70 & 0.028 & 0.1 & 
16.9$\times 10^6$\\
3:$\times$ & 4.064 & 4.347 & 3.725 & 10.23 & 
  7.27 & 0.249 & 0.70 & 0.020 & 
0.2 & 17.9$\times 10^6$\\
4:$\times$ & 4.026 & 4.339 & 3.721 & 10.04 & 7.23 & 0.255 & 0.70 & 0.022 & 
0.2 & 18.6$\times 10^6$\\
5:$\times$ & 3.992 & 4.331 & 3.718 & 9.88 & 7.20 & 0.260 & 0.70 & 0.024 & 
0.2 & 19.3$\times 10^6$\\
\enddata\label{table2}
\end{deluxetable}

\end{document}